\documentclass{hosotani}

\def\dd{\partial}

\def\la{\raise.16ex\hbox{$\langle$} \, }
\def\ra{\, \raise.16ex\hbox{$\rangle$} }
\def\go{\rightarrow}

\begin{document}

\title{COMPLEX MONOPOLES IN THE PATH INTEGRAL\footnote{To appear in
the Proceedings for {\it the 6th International Conference on PATH-INTEGRALS
from peV to TeV}, Florence, Italy, 25-29 August 1998.}}

\author{Y.\ Hosotani and B.\ Tekin}

\address{School of Physics and Astronomy, University of Minnesota\\
Minneapolis, MN 55455, USA}

\author{K.\ Saririan}

\address{Infinity, A Sungard Company\\
 640 Clyde Court, Mountain View, CA 94043, USA}  


\maketitle

\abstracts{Complex monopole configurations dominate in the path integral
in the Georgi-Glashow-Chern-Simons  model and disorder the Higgs vacuum.
No cancellation is expected among Gribov copies of the  monopole
configurations.}

\centerline{\small ( hep-th/9809072 ~ UMN-TH-1718/98 )}

\section{Georgi-Glashow-Chern-Simons model}

The Georgi-Glashow model is a $SO(3)$ gauge theory with a triplet Higgs
scalar field $\vec h$ in which the gauge symmetry is
spontaneously broken to $U(1)$ by the Higgs mechanism.   The vacuum
is ordered with   nonvanishing $\la \vec h \ra \not= 0$.

In three dimensions instantons, or monopoles, disorder the Higgs vacuum;
$\la \vec h \ra =0$.   Electric charges are linearly confined, forming an
electric flux  string.\cite{Polyakov}   The model is dual to the Josephson
junction  system in the superconductivity.\cite{Hosotani}

Further the Chern-Simons term can be added 
to the Lagrangian.   This defines the Georgi-Glashow-Chern-Simons  model.
The $U(1)$ gauge boson acquires
a topological mass, and electric charges are screened.   

How about the Higgs vacuum?  Is the vacuum still disordered
such that $\la \vec h \ra =0$?  In disordering the vacuum, monopole
configurations play an important role. It has been argued  in
the literature,\cite{D'Hoker}  however, that
monopole configurations  would become irrelevant once the Chern-Simons
term is added;    monopole solutions would have infinite action, and for
configurations of finite action their Gribov copies would lead to
cancellation. 
We are going to show that this is not the case.  There are complex
monopole solutions of finite action, and Gribov copies do not lead to
cancellation.\cite{Tekin}

\section{Monopole ansatz}

The most general form of the spherically symmetric
monopole ansatz is 
\bea
&&h^a(\vec{x})=\hat x^a h(r) \cr
&&A^a_\mu(\vec{x})= {1\over r} \left[ \epsilon_{a\mu
\nu}\hat{x}^\nu(1-\phi_1) 
+ ( \delta_{a\mu} - \hat{x}_a  \hat{x}_\mu) \phi_2 
    + rS \hat{x}_a  \hat{x}_\mu\right]       
\label{configuration1}       
\eea 
where $\hat x^a = x^a/r$.  
The regularity of configurations at the origin and the finiteness of the 
action impose boundary conditions $(h,\phi_1,\phi_2)=(0,1,0)$ at $r=0$
and $(h,\phi_1,\phi_2,S)=(v,0,0,0)$ at $r=\infty$.

Under a gauge transformation $A \go \Omega A \Omega^{-1} + \Omega d
\Omega^{-1}$ where
$\Omega=\exp \big\{ {i\over 2} f(r) \hat x^a \sigma^a \big\} $ and 
 $f(0)=0$,
\be
\pmatrix{\phi_1\cr \phi_2\cr} \go 
\pmatrix{\cos f & \sin f \cr -\sin f & \cos f \cr} 
  \pmatrix{\phi_1\cr \phi_2\cr} ~~,~~
S ~~ \go ~~ S - f' ~~.
\label{transformation1}
\ee
The Chern-Simons term, $I_{CS} = -(i\kappa/g^2)\int 
~ \hbox{tr}\left( A \wedge d A +{1\over 3} A \wedge A \wedge A \right)$,
 is not gauge invariant;
$\delta I_{CS} = (4\pi i\kappa/g^2) f(\infty)$.
On $S^3$ $f(\infty)$ is a multiple of $2\pi$ so that the
quantized Chern-Simons coefficient guarantees the gauge invariance.
On $R^3$, however, there is a priori no reason to demand that $f(\infty)$
be quantized.

\section{Path integral and complex monopoles}

In the path integral the gauge fixing condition is inserted;
\be
Z = \int {\cal D}A^a_\mu {\cal D}\vec h 
~ \Delta_{FP}[A] ~\delta[F(A)] ~ e^{-I} ~.
\label{PI}
\ee
We look for configurations which extremize the action $I$ within
the subspace specified with $F(A)=0$.  

In the radial gauge  $S=0$ the
extremization of the action leads to
\bea
&&\phi_1'' + {1\over r^2} (1-\phi_1^2  -\phi_2^2) \phi_1
 + i\kappa \phi_2' - h^2 \phi_1 = 0 \label{EqMotion6} \cr
&&\phi_2'' + {1\over r^2} (1-\phi_1^2  -\phi_2^2) \phi_2
 - i\kappa \phi_1' - h^2 \phi_2 = 0  \cr
&&{1\over r^2} {d\over dr} \Big( r^2 {dh\over dr}  \Big) 
- \lambda (h^2 - v^2) h -
{2\over r^2} (\phi_1^2 + \phi_2^2) h = 0 ~~.
\label{EqMotion1}
\eea
Since eq.\ (\ref{EqMotion1}) contains complex terms,
solutions necessarily become complex. 

\begin{figure}[t]\centering
\mbox{
\epsfysize=5cm \epsfbox{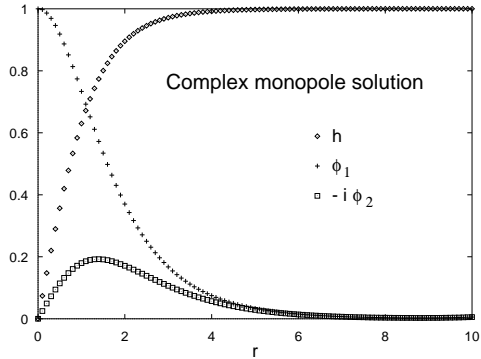}}
\vskip -.3cm
\caption{Complex monopole solution in the radial gauge for $v=1$, $\kappa=.5$, 
and $\lambda=.5$.  $\phi_2(r)$ is pure imaginary.} 
\end{figure}

Eq.\ (\ref{EqMotion1})  is solved by an ansatz
$\phi_1 = \zeta(r) \cosh {1\over 2} \kappa r$ and $\phi_2 = i \zeta(r)
\sinh {1\over 2} \kappa r$.  The solution is depicted in fig.\ 1.
$\phi_2(r)$ is pure imaginary.  The action is real and finite.  The
$U(1)$ field strengths are given  exactly by those of a real magnetic
monopole.  Non-Abelian field strengths are complex.   There is no Gribov
copy in this gauge.

In the original form 
of the path integral, field configurations are integrated along real
axes.  We have found that   the saddle 
points of $I[A,h]$  are located off the real axes.  In the 
saddle point method for the integration, the integration path is
deformed such that a new path pass the saddle points.  The complex
monopole configurations approximate the integral, and  dominate the path
integral.  They are relevant in disordering the Higgs vacuum.  Without
monopole-type configurations the perturbative Higgs vacuum cannot
be disordered and $\la \vec h \ra$ remains nonvanishing.  With
complex monopoles taken into account $\la \vec h \ra =0$ but $\la
\vec h^2 \ra \sim v^2$.

We remark that if the gauge is not fixed and the action is varied with
respect to arbitrary gauge field configurations, then one would obtain one
more equation to be solved.  This equation is not satisfied by our
solution.  But in the path integral the configuration space is restricted
by the gauge condition as in (\ref{PI}).  This subtlety arises due to the 
gauge non-invariance of the Chern-Simons term.

\section{Gribov copies}

The radiation gauge does not uniquely fix gauge field
configurations.\cite{Gribov}  In the monopole ansatz
the radiation gauge
condition $\dd_\mu A^a_\mu =0$ is maintained if  $f(r)$ in
(\ref{transformation1}) obeys
$f'' + (2/ r) f' - (2/ r^2) \big\{ \phi_1 \sin f
+ \phi_2 (1-\cos f) \big\} = 0$.
 Solutions to this equation define Gribov copies.  

These copies have a significant effect in the Chern-Simons theory.
The Chern-Simons term is not gauge invariant.  Gribov copies carry
an extra phase factor, $\exp \big\{ (4\pi i\kappa/g^2) f(\infty) \big\}$,
which could lead to cancellation in the path integral.

\begin{figure}[t]\centering
\mbox{
\epsfysize=5cm \epsfbox{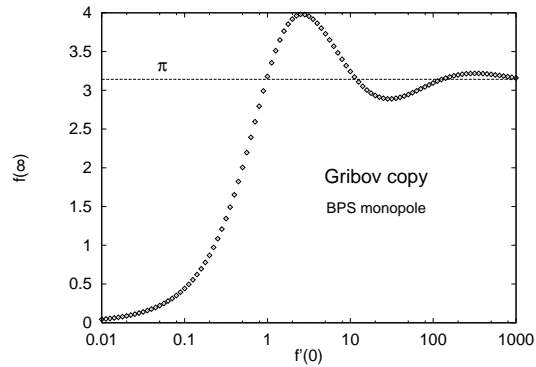}}
\vskip -.3cm
\caption{$f'(0)$ vs $f(\infty)$ for Gribov copies of the BPS monopole.} 
\end{figure}

Solutions $f(r)$ are uniquely determined by $f(0)=0$ and
$f'(0)$.  In fig.\ 2 we have plotted $f(\infty)$ as a function of $f'(0)$
for the BPS monopole solution. The 
range of the asymptotic value is $-3.98 < f(\infty) < + 3.98$.  It is
quite unlikely that  these
Gribov copies of the BPS monopole lead to the cancellation $\sum
e^{-4\pi i\kappa f(\infty)/g^2} =0$ in the presence of the Chern-Simons
term. Monopole configurations remain important in the path integral.

\section*{Acknowledgments}
We would like to thank R.\ Jackiw for his enlightening comments 
in the conference.   This work was
supported in part  by  the U.S.\ Department of Energy under contracts
DE-FG02-94ER-40823.

\def\AP{{\em Ann.\ Phys.\ (N.Y.)} }


\begin{thebibliography}{99}
\bibitem{Polyakov}
A.\ M.\ Polyakov, \Journal{\PLB}{59}{80}{1975};
\Journal{\NPB}{120}{429}{1977}. 

\bibitem{Hosotani}
Y.\ Hosotani, \Journal{\PLB}{69}{499}{1977}.


\bibitem{D'Hoker} 
E.\ D.\ D'Hoker and L.\ Vinet, \Journal{\AP}{162}{413}{1985};
R.\ D.\ Pisarski, \Journal{\PRD}{34}{3851}{1986}; 
I.\ Affleck, J.\ Harvey, L.\ Palla and G.\ Semenoff, 
\Journal{\NPB}{328}{575}{1989};
E.\ Fradkin and F.\ A.\ Schaposnik, \Journal{\PRL}{66}{276}{1991};
K.\ Lee, \Journal{\NPB}{373}{735}{1992};
\Journal{\NPB}{425}{137}{1994};
R.\ Jackiw and S.Y.\  Pi, \Journal{\PLB}{423}{364}{1998};

\bibitem{Tekin} B.\ Tekin, K.\ Saririan, and Y.\ Hosotani,
preprints hep-th/9808045; hep-th/9808057; hep-th/9808105.

\bibitem{Gribov} V.\ N.\ Gribov, \Journal{\NPB}{139}{1}{1978}.

\end{thebibliography}
\end{document}